\documentclass[sigconf, screen]{acmart}
\setcopyright{rightsretained}
\acmPrice{}
\acmDOI{10.1145/3540250.3560881}
\acmYear{2022}
\copyrightyear{2022}
\acmSubmissionID{fse22ivr-p19-p}
\acmISBN{978-1-4503-9413-0/22/11}
\acmConference[ESEC/FSE '22]{Proceedings of the 30th ACM Joint European Software Engineering Conference and Symposium on the Foundations of Software Engineering}{November 14--18, 2022}{Singapore, Singapore}
\acmBooktitle{Proceedings of the 30th ACM Joint European Software Engineering Conference and Symposium on the Foundations of Software Engineering (ESEC/FSE '22), November 14--18, 2022, Singapore, Singapore}

\usepackage[title]{appendix}

\usepackage{amsmath,amsfonts}
\usepackage{algorithmic}
\usepackage{graphicx}
\usepackage{textcomp}
\usepackage{xcolor}
\usepackage{booktabs} % For formal tables \toprule
\usepackage{xspace} % Put an \xspace within numeric macros
\usepackage[normalem]{ulem}
\usepackage{makecell}
\usepackage{tcolorbox}
\usepackage{enumitem}
\usepackage{siunitx}
\usepackage[vskip=1em,font=itshape,leftmargin=2em,rightmargin=2em]{quoting}
\usepackage{soul}

\usepackage{cleveref}

\crefformat{section}{\S#2#1#3}
\crefname{figure}{Figure}{Figures}
\crefname{appendix}{Appendix}{Appendices}
\crefname{table}{Table}{Tables}
\crefname{algorithm}{Algorithm}{Algorithms}
\crefname{listing}{Listing}{Listings}
\crefname{theorem}{Theorem}{Theorems}
\crefname{thm}{Theorem}{Theorems}
\crefname{lemma}{Lemma}{Lemmata}
\crefname{equation}{Eqt.}{Eqts.}
\crefformat{Grammar}{Grammar #1}

    \usepackage{etoolbox}
    \makeatletter
    \patchcmd{\ttlh@hang}{\parindent\z@}{\parindent\z@\leavevmode}{}{}
    \patchcmd{\ttlh@hang}{\noindent}{}{}{}
    \makeatother

\newif\ifSPACEHACK
\SPACEHACKfalse

\ifSPACEHACK
    \usepackage{titlesec}
    \titleformat{\section}
      {\normalfont\Large\bfseries}{\thesection}{1em}{\MakeUppercase}
    \titlespacing*\section{0pt}{1pt plus 1pt minus 1pt}{1pt plus 1.5pt minus 1.5pt}
    \titlespacing*\subsection{0pt}{1pt plus 1.5pt minus 1.5pt}{1pt plus 1.5pt minus 1.5pt}
    \titlespacing*\subsubsection{0pt}{2pt plus 1pt minus 1pt}{1pt plus 1.5pt minus 1.5pt}
    \titlespacing*\paragraph{0pt}{1pt plus 1.5pt minus 1.5pt}{1pt plus 1.5pt minus 1.5pt}
\else
\fi

\newcommand{\eg}{\textit{e.g.,} }
\newcommand{\etal}{\textit{et al.}\xspace}

\begin{document}

\title[Discrepancies among Pre-trained Deep Neural Networks: A New Threat to Model Zoo Reliability]{Discrepancies among Pre-trained Deep Neural Networks:\\A New Threat to Model Zoo Reliability}

\author{Diego Montes}
\orcid{0000-0001-7994-1281}
\affiliation{%
  \institution{Purdue University}
  \country{USA}
}
\email{montes10@purdue.edu}

\author{Pongpatapee Peerapatanapokin}
\orcid{0000-0002-8901-7495}
\affiliation{%
  \institution{Purdue University}
  \country{USA}
}
\email{ppeerapa@purdue.edu}

\author{Jeff Schultz}
\orcid{0000-0002-1470-7897}
\affiliation{%
  \institution{Purdue University}
  \country{USA}
}
\email{schul203@purdue.edu}

\author{Chengjun Guo}
\orcid{0000-0002-6314-2028}
\affiliation{%
  \institution{Purdue University}
  \country{USA}
}
\email{guo456@purdue.edu}

\author{Wenxin Jiang}
\orcid{0000-0003-2608-8576}
\affiliation{%
  \institution{Purdue University}
  \country{USA}
}
\email{jiang784@purdue.edu}

\author{James C. Davis}
\orcid{0000-0003-2495-686X}
\affiliation{%
  \institution{Purdue University}
  \country{USA}
}
\email{davisjam@purdue.edu}

\begin{abstract}

Training deep neural networks (DNNs) takes significant time and resources. A practice for expedited deployment is to use pre-trained deep neural networks (PTNNs), often from model zoos---collections of PTNNs; yet, the reliability of model zoos remains unexamined. In the absence of an industry standard for the implementation and performance of PTNNs, engineers cannot confidently incorporate them into production systems. 
As a first step, discovering potential discrepancies between PTNNs across model zoos would reveal a threat to model zoo reliability.
Prior works indicated existing variances in deep learning systems in terms of accuracy.
However, broader measures of reliability for PTNNs from model zoos are unexplored. This work measures notable discrepancies between accuracy, latency, and architecture of 36 PTNNs across four model zoos.
Among the top 10 discrepancies, we find differences of 1.23\%--2.62\% in accuracy and 9\%--131\% in latency. We also find mismatches in architecture for well-known DNN architectures (\eg ResNet and AlexNet).
Our findings call for future works on empirical validation, automated tools for measurement, and best practices for implementation.

\end{abstract}

\begin{CCSXML}
<ccs2012>
   <concept>
       <concept_id>10011007.10011074.10011092.10011096</concept_id>
       <concept_desc>Software and its engineering~Reusability</concept_desc>
       <concept_significance>500</concept_significance>
       </concept>
   <concept>
       <concept_id>10010147.10010257.10010293.10010294</concept_id>
       <concept_desc>Computing methodologies~Neural networks</concept_desc>
       <concept_significance>500</concept_significance>
       </concept>
 </ccs2012>
\end{CCSXML}



\maketitle

\vspace{-2mm}
\section{Introduction}

With the growing energy consumption of training DNNs~\cite{patterson2021carbon}, taking advantage of the re-usability of PTNNs can significantly reduce the costs of training~\cite{han}. In particular, transfer learning can result in shorter training times and higher asymptotic accuracies compared to other weight initialization methods~\cite{liuhpc, Thrun1998}. This kind of technique accelerates model reuse and development. The history of PTNNs and their impact on the development of artificial intelligence has been extensively documented~\cite{han, Pan}. As such, collections of PTNNs have been created, referred to as \emph{model zoos}. Notably, maintainers of popular machine learning frameworks, such as TensorFlow~\cite{tensorflow2015-whitepaper}, maintain corresponding model zoos developed with their framework, such as the TensorFlow Model Garden~\cite{tensorflowmodelgarden2020}.

There are many model zoos~\cite{torchvision, tensorflowmodelgarden2020, KerasApplication, ONNX}
and an expanding use of PTNNs in production systems~\cite{han}. Past work has emphasized the difficulties in adopting software engineering practices in machine learning, and specifically, the challenges with reproducing machine learning research papers~\cite{banna2021experience, reproducescience}. These reproducibility issues may affect PTNNs, leading to variations across model zoos~\cite{pham}.
Disparities in the accuracy, latency, or architecture of a PTNN could negatively affect a deep learning system, threatening PTNNs' reuse potential. Consider a model zoo that has an incorrect implementation of a well-known DNN architecture, which has increased its latency significantly. If an engineer were to use the PTNN from this zoo, they would unknowingly be receiving a lower quality PTNN than they might otherwise have from a different model zoo. The engineer's effort to enable a quick turnaround time with a PTNN would have become harmful. Discovering discrepancies would shine a light on the reliability of model zoos.

To explore the reliability of model zoos, we performed a measurement study to identify discrepancies among 36 image classification PTNN architectures across four model zoos: \emph{TensorFlow Model Garden} (TFMG)~\cite{tensorflowmodelgarden2020}, \emph{ONNX Model Zoo} (ONNX)~\cite{ONNX}, \emph{Torchvision Models} (Torchvision)~\cite{torchvision}, and \emph{Keras Applications} (Keras)~\cite{KerasApplication}. The PTNNs were measured along three dimensions: accuracy, latency, and architecture. We find the differences in accuracies on \emph{ILSVRC-2012-CLS} dataset (ImageNet) can be as large as 2.62\% ~\cite{deng2009imagenet}.\footnote{The \emph{ILSVRC-2012-CLS} image dataset has 50,000 validation images. A 1\% accuracy difference is equivalent to 500 images.} Similarly, over 20\% of the PTNNs measured had latency differences (FLOPs) of 10\% or more when comparing PTNNs of the same name across the model zoos. Lastly, we discover architectural differences in several PTNNs, including implementations of \emph{AlexNet} and \emph{ResNet V2}. We conclude with an agenda for future research on further empirical validation, automated tools for measurement, and best practices for implementing model zoo PTNNs.

\section{Background and Related Work}

PTNNs are applied in a wide variety of domains~\cite{han}.
With the demand for engineers far exceeding supply~\cite{Sakurada2021Industry4NewJob}, companies are looking for best practices that can boost the productivity of their engineers. 
Major companies (\eg Google and Microsoft) have shared best practices on machine learning development and informed future directions on model reuse~\cite{Breck2017aRubricforMLProductionReadinessandTechnicalDebtReduction, 8804457}.
A case study from SAP indicates possible compatibility, portability, and scalability challenges in machine learning model deployment, which may affect their performance~\cite{Rahman2019MLSEinPractice}. 
There have been many efforts to improve the quality of model zoos. 
For example, IBM has developed a tool to extract model metadata~\cite{IBM2020AIMMX} to support better model management.
Banna \etal promote best practices for reproducing and publishing high-quality PTNNs~\cite{banna2021experience}. 
However, the reliability of model zoos has not been validated by prior works.

The ability to replicate the accuracy of a DNN in identical training environments is hindered by non-deterministic factors. Accuracy differences of up to 10.8\%, stemming purely from non-determinism, have been reported with popular DNN architectures \cite{pham}. Closely related, research has investigated and benchmarked the performance variances tied to deep learning frameworks~\cite{park, liu}. This variability threatens the reliability of new deep learning techniques. As such, automated variance testing~\cite{phamdeviate} has been proposed to assure the validity of these comparisons. 
However, PTNNs in model zoos may also suffer from varying architectural implementations, affecting more than just accuracy. Our work measures the disparities in PTNNs across different model zoos as opposed to attempting to improve the standard in just one~\cite{banna2021experience}. Our results enlighten future works validating the quality and promoting the standardization of model zoos.

\section{Methodology}

We perform a measurement study to assess our problem statement: whether discrepancies exist between the accuracy, latency, and architecture of PTNNs across different model zoos.

\subsection{Subjects}

A \emph{model zoo} is a collection of PTNNs for various tasks. We carry out a selection process for four model zoos. Our selection criteria included the model zoo being maintained alongside a machine learning framework: this increases the likelihood of the model zoo being actively maintained. Furthermore, to ensure the popularity of the model zoo, the zoo must have a public GitHub repository with at least three thousand stars~\cite{whatsinastar}. 
Using GitHub search\footnote{https://github.com/search} to identify potential model zoo candidates, 11 model zoos were selected that met the criteria.\footnote{The 11 identified potential model zoos are as follows: TensorFlow Model Garden, ONNX Model Zoo, Torchvision Models, Keras Applications, TensorFlow Model Hub, PyTorch Model Zoo, MXNet Model Zoo, Gluon Model Zoo, Deeplearning4j Model Zoo, Caffe Model Zoo, and OpenVINO Model Zoo.} The PTNNs within the 11 model zoos were categorized into deep learning tasks, including image classification, object detection, and natural language processing. We focused on image classification models because it is the most common type in 8 of the 11 model zoos.

A PTNN availability analysis was done on the candidate model zoos to assess how many model zoos offered the same image classification PTNN architectures. Based on the largest shared availability, we chose \emph{TensorFlow Model Garden}, \emph{ONNX Model Zoo}, \emph{Torchvision Models}, and \emph{Keras Applications}. Within these model zoos, we selected all the image classification PTNN architectures that were present in at least two of the four model zoos, yielding 36 PTNN architectures. The selected PTNNs are either directly downloadable from the model zoos' GitHub repositories or can be pulled using the model zoos' APIs.

{
\small
\renewcommand{\arraystretch}{1}
\begin{figure}
    \centering
    \includegraphics[scale=0.55]{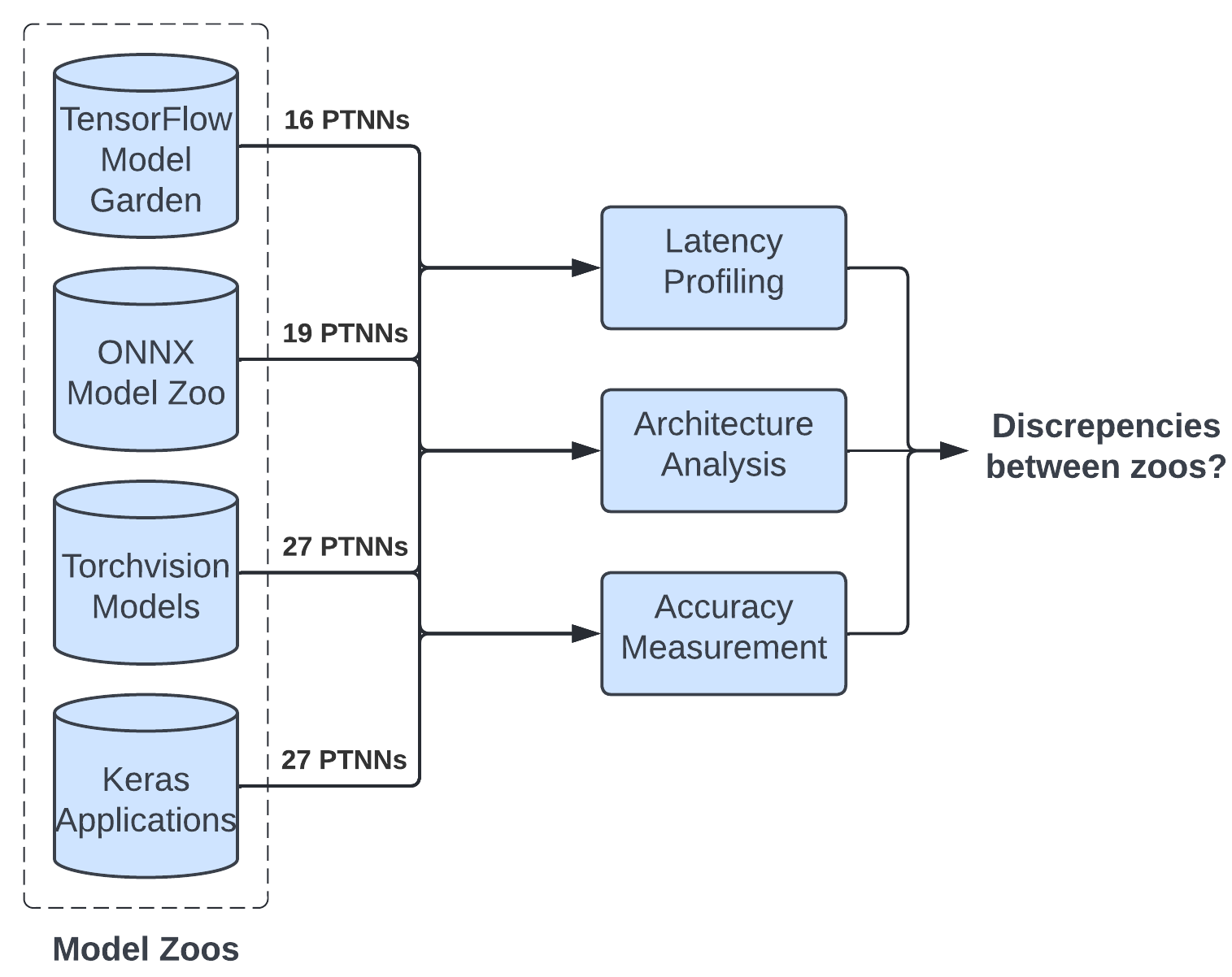}
    \vspace{-4mm}
    \caption{Overview of the measurement process. We gather PTNNs from the model zoos with the same name, perform measurements on each PTNN, and compare for discrepancies.}
    \label{fig:Overview of the measurement process}
    \vspace{-5mm}
\end{figure}
}

\subsection{Evaluation Metrics}
\textbf{Accuracy.} Image classification DNNs' effectiveness is measured in accuracy, which is a critical component of a PTNN. We are measuring discrepancies between the claims of model zoos as opposed to verifying them. \textit{Top-1 accuracy} is the conventional accuracy where model prediction must exactly match the expected label, while \textit{top-5 accuracy} measures the fraction of images where any of the top five predicted labels matches the target label~\cite{Dang2021AccuracyandLoss, deng2009imagenet}. 35 image classification PTNN architectures reported top-1 ImageNet classification accuracies, meanwhile only 32 reported top-5 ImageNet classification accuracies.

\noindent\textbf{Latency.} The latency of a DNN is a key factor that engineers consider~\cite{gopalakrishna_iot}. For example, \emph{MobileNet} is a DNN image classification architecture that prioritizes low latency on mobile and embedded systems~\cite{DBLP:journals/corr/HowardZCKWWAA17}. 
We used open-source tools~\cite{onnxopcounter, fvcore, tensorflowrepo} to measure the latency by counting the floating point operations (FLOPs)~\cite{flopsbench}.
FLOPs are framework and hardware-agnostic, allowing for unbiased comparisons. 

\noindent\textbf{Architecture.} PTNNs are trained weights based on research papers that propose DNN architectures. As a result, model zoos advertise PTNNs by their architecture name. The observed accuracy differences and past work on DNN vulnerabilities motivated us to examine architecture~\cite{gu2019badnets}. Qualitative observations of discrepancies in the descriptions, source code, and visualizations of PTNN architectures were employed. Specifically, netron, an open-source neural network visualizer, was used to inspect the architecture of the PTNNs~\cite{netron}. However, not all neural network weight formats are supported, so all PTNNs were converted to the ONNX format for architectural analysis using an appropriate tool for each framework~\cite{tf2onnx, pytorch}. The source code for the implementations of the PTNNs are present in the model zoos' GitHub repositories and was used as an additional form of PTNN inspection.

\section{Results and Analysis}

\subsection{Accuracy}

We compared the top-1 accuracy of 35 PTNN architectures and the top-5 accuracy of 32 PTNN architectures by using ImageNet. Notably, 12 of the 35 profiled PTNN architectures had top-1 accuracy differences greater than 0.96\%. For top-5 accuracies, 6 of the 32 PTNN architectures had differences greater than 0.94\%. The large differences present in Figure \ref{fig:Top 10 Largest Top-1 Accuracy Differences} have significant consequences. For example, \emph{ResNet V1 152} from Keras is noticeably less accurate than the PTNN by the same name from Torchvision, with top-1 accuracies of 76.6\% and 78.31\%, respectively. This difference is pronounced enough that \emph{ResNet V1 101} from Torchvsion with top-1 accuracy of 77.37\% is more accurate than \emph{ResNet V1 152} from Keras.\footnote{ResNet V1 101 was originally reported to be 0.32\% less accurate than \emph{ResNet V1 152}~\cite{7780459}.}

{
\small
\renewcommand{\arraystretch}{0.9}
\vspace{-4mm}
\begin{figure}[h]
    \centering
    \includegraphics[scale=0.26]{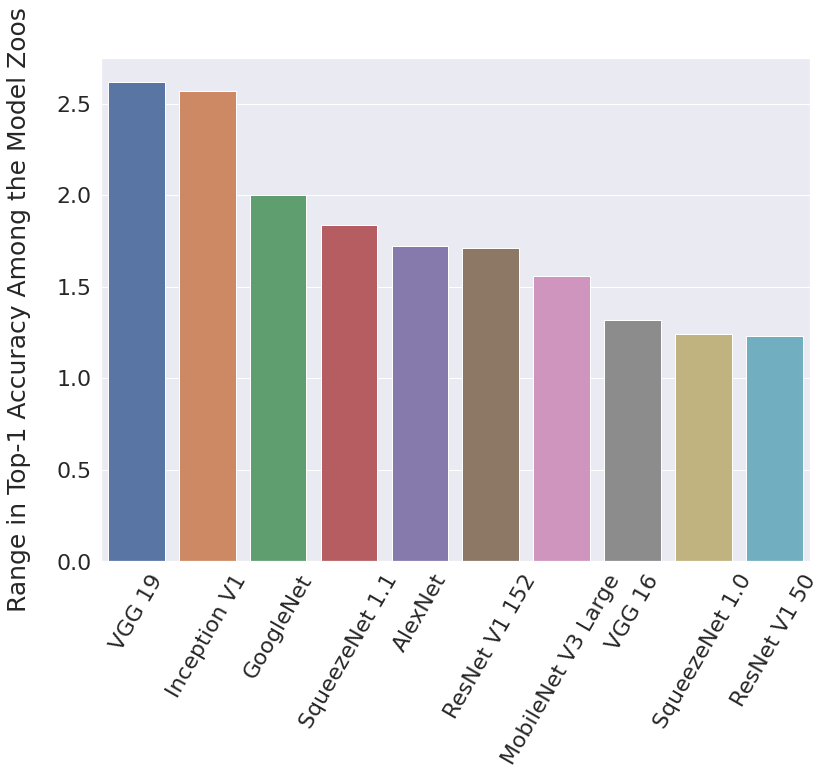}
    \vspace{-4mm}
    \caption{Top 10 largest top-1 accuracy differences. For a PTNN architecture, the accuracy of the PTNN with the lowest reported top-1 accuracy is subtracted from that of the PTNN with the largest top-1 accuracy.}
    \label{fig:Top 10 Largest Top-1 Accuracy Differences}
    \vspace{-3mm}
\end{figure}
}

Table \ref{Frequency accuracy table} shows the aggregation of accuracy differences across model zoos, highlighting how often a model zoo had the highest or lowest top-1 or top-5 accuracy for a given PTNN architecture. As seen, 48\% of the PTNNs that were available on Torchvision had the highest top-1 accuracy among the model zoos. On the other hand, Keras had the lowest top-1 accuracy 44\% of the time for its selection of PTNNs.

\begin{table}[t]
    \vspace{-3mm}
    \caption{Frequency at which each model zoo had the most or least accurate model ordered by highest top-1 accuracy.}
\resizebox{\columnwidth}{!}{    \begin{tabular}{lcccc}
      & Highest Top-1 & Lowest Top-1 & Highest Top-5 & Lowest Top-5 \\
      \midrule
      Torchvision Models & 48\% & 41\% & 52\% & 36\% \\
      TF Model Garden & 40\% & 33\% & 36\% & 43\%\\
      Keras Applications & 37\% & 44\% & 36\% & 40\% \\
      ONNX Model Zoo & 35\% & 41\% & 31\% & 44\% \\
      \bottomrule
    \end{tabular}}
    \label{Frequency accuracy table}
    \vspace{-3mm}
\end{table}

\subsection{Latency}

36 PTNN architectures were measured for their FLOPs. Figure \ref{fig:Top 10 Largest FLOPs Differences} shows that there are 8 PTNN architectures where the PTNN with the highest amount of FLOPs had greater than 10\% more FLOPs than the PTNN with the lowest FLOP count. At the extreme, Torchvision's \emph{SqueezeNet 1.0}, sitting at 819.08 million FLOPs, had 2.31$\times$ the FLOPs of ONNX's \emph{SqueezeNet 1.0}. Likewise, the three PTNN architectures from the \emph{ResNet V2} family all had greater than 85\% more FLOPs than the lowest FLOPs PTNN. All the high FLOP-count \emph{ResNet V2} come from TFMG.

We discuss the possible explanations for the FLOPs differences seen in Figure \ref{fig:Top 10 Largest FLOPs Differences}. The high FLOPs difference measured in \emph{SqueezeNet 1.0} can be explained by looking at its successor, \emph{SqueezeNet 1.1}. \emph{SqueezeNet 1.1} is advertised by ONNX to contain 2.4$\times$ less computation than the former. However, \emph{SqueezeNet 1.1} from ONNX has the same number of measured FLOPs as the \emph{1.0} PTNN offered. ONNX has been advertising \emph{SqueezeNet 1.1} as its \emph{1.0} counterpart. Similarly, looking at the \emph{ResNet V2} from TFMG: a primary contributor to the large amount of FLOPs is the input shape. \emph{ResNet V2} architectures, according to the origin paper, accept 224$\times$224 inputs \cite{10.1007/978-3-319-46493-0_38}; however, TFMG states that the \emph{ResNet V2} PTNNs it provides use Inception pre-processing and an input image size of 299$\times$299. A trade-off between accuracy and throughput, FLOPs, was potentially made here by the model zoo maintainers.

Across all FLOP-counted PTNNs, Torchvision had the highest FLOPs PTNNs for 78\% of the PTNNs it offered. Close behind, TFMG had 69\%. Pointedly, Keras never had the highest FLOPs PTNN and had the lowest FLOPs implementation 81\% of the time.

{
\small
\renewcommand{\arraystretch}{0.9}
\begin{figure}
    \centering
    \includegraphics[scale=0.26]{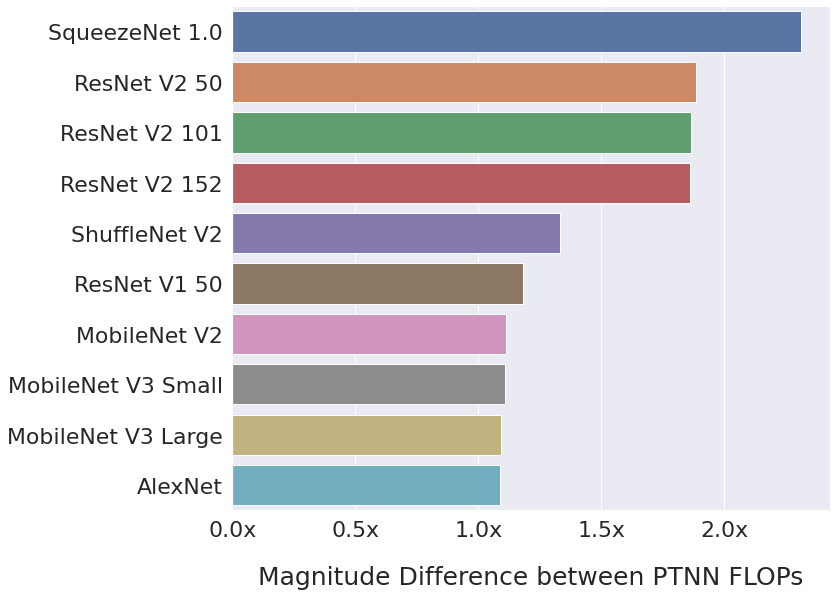}
    \vspace{-4mm}
    \caption{Top 10 largest FLOPs differences. For a PTNN architecture, the FLOP count of the PTNN with the most FLOPs is divided by the FLOPs of the PTNN with the fewest.}
    \label{fig:Top 10 Largest FLOPs Differences}
    \vspace{-5mm}
\end{figure}
}

\subsection{Architecture} \label{Architecture}

We frame our results for architecture in terms of the discrepancies we discovered in our analysis. Specifically, we discuss differences among PTNNs for \emph{AlexNet}, \emph{ResNet V1 101}, \emph{ResNet V2 50}, and \emph{ResNet V2 101} and against the PTNNs' origin papers.

The \emph{AlexNet} from Torchvision cites a different origin paper than other model zoos \cite{DBLP:journals/corr/Krizhevsky14, 10.5555/2999134.2999257}. Both papers contain the same first author; however, only the latter contains an explicit description of a DNN architecture. As such, our analysis pegs the PTNN against the latter paper \cite{10.5555/2999134.2999257}. We notice two main discrepancies: the PTNN is missing the response normalization layers and the kernel-size and number of kernels for the convolution layers are incorrect. For instance, Torchvision's PTNN has 64 kernels in the first convolution layer as opposed to the 96 that are described in the origin paper.

The \emph{ResNet V1 101} from ONNX and Keras contain convolution shortcuts, which were only introduced in the \emph{ResNet V2} paper, but not in the \emph{ResNet V1} origin paper \cite{10.1007/978-3-319-46493-0_38, 7780459}. Torchvision's and TFMG's \emph{ResNet V1 101} do not include this shortcut. Also in the \emph{ResNet} family, both the \emph{ResNet V2 50} and \emph{ResNet V2 101} have a shared discrepancy. As seen in Figure \ref{fig:ResNet V2 50 architecture}, Keras' \emph{ResNet V2 50} implementation contains max pool skip connections, which are not present in the paper, and uses convolutions with larger strides in these residual blocks \cite{10.1007/978-3-319-46493-0_38}.

The observed discrepancies in architecture may affect the accuracy and latency. 
For example, the larger convolution strides and max pool skip connection in the \emph{ResNet V2 50} from Keras allows the network to use less compute, FLOPs, compared to the PTNN from ONNX. This can be seen in the FLOP measurements of the \emph{ResNet V2 50} from Keras and ONNX. ONNX's \emph{ResNet V2 50} has 4.12 billion FLOPs while Keras' PTNN only has 3.49 billion FLOPs, an 18.1\% difference. 
Moreover, the Keras PTNN did not sacrifice accuracy through this implementation, reporting a 76\% top-1 accuracy, which is greater than ONNX's \emph{ResNet V2 50} top-1 accuracy of 75.81\%.  While the Keras maintainers did not implement \emph{ResNet V2 50} faithfully to the origin paper, they produced a more accurate PTNN with lower latency.

{
\small
\renewcommand{\arraystretch}{0.7}
\begin{figure}
    \centering
    \includegraphics[scale=0.22]{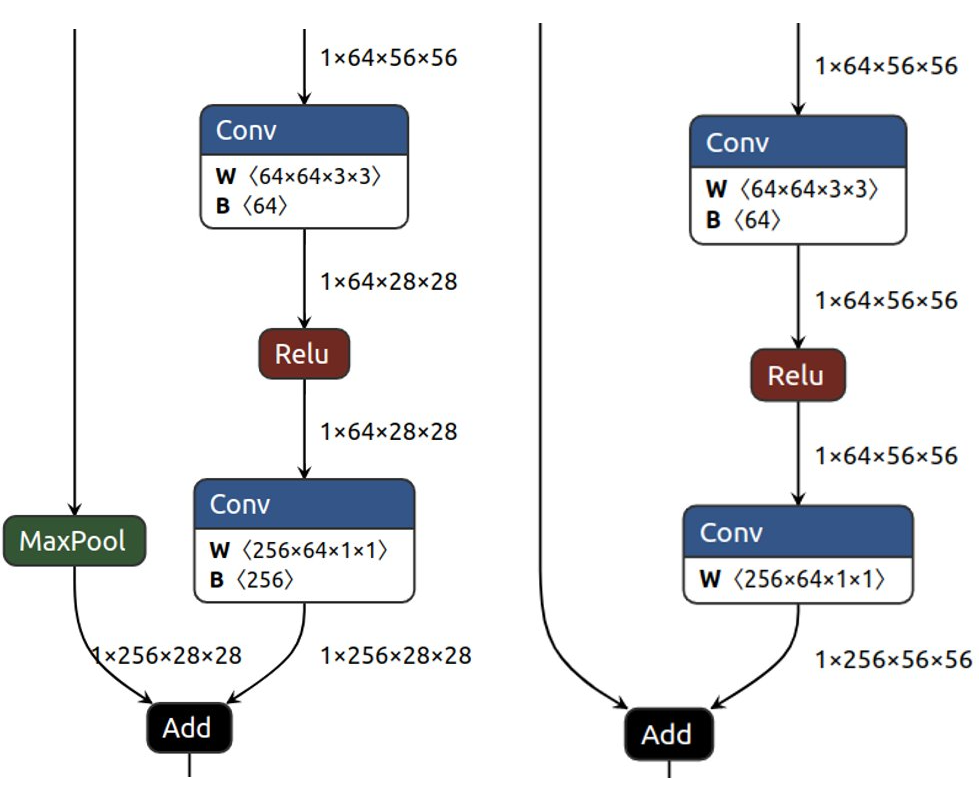}
    \vspace{-3mm}
    \caption{\emph{ResNet V2 50} architecture differences between \emph{Keras Applications} (left) and \emph{ONNX Model Zoo} (right). The top-right convolution on the left has a stride size of 2, while the top-right convolution on the right has a stride size of 1.}
    \label{fig:ResNet V2 50 architecture}
    \vspace{-5mm}
\end{figure}
}

\section{Discussion and Future Work}

\noindent\textbf{Empirical Validation.} The top-1 accuracy differences depicted in Figure \ref{fig:Top 10 Largest Top-1 Accuracy Differences} suggest that the choice of model zoo matters. Specifically, 34\% of the PTNN architectures having top-1 accuracy differences greater than 0.96\% is not easily overlooked. An engineer may receive a PTNN that incorrectly classifies greater than 500 validation images on ImageNet more than a PTNN from a different model zoo. Model zoo choice should not result in a noticeable impact on the accuracy of PTNNs that engineers receive. Although model zoos currently report the accuracy of the PTNNs they offer, our work has shown that this does not guarantee that there is not another model zoo with the same PTNN at a higher accuracy. Publicly available and actively maintained comparisons of model zoo PTNNs would allow engineers to be more informed when choosing a model zoo. Furthermore, we only studied the accuracies of image classification models at face value. 
We recommend future works focus on empirical validation on the claims of PTNNs in model zoos to check for the existence of false advertising. 

\noindent\textbf{New Metrics and Automated Tools.} The measured FLOP disparities seen in \cref{fig:Top 10 Largest FLOPs Differences} have consequences, especially in edge devices with limited compute. For example, ONNX incorrectly listing \emph{SqueezeNet 1.1} as \emph{SqueezeNet 1.0} may lead to confusion when an engineer switches to \emph{SqueezeNet 1.1} from \emph{SqueezeNet 1.0} expecting a drop in latency. Similar confusion may arise from instances like the one seen in TFMG's selection of \emph{ResNet V2}. While the increased input size is stated, the impact on latency is not made clear. To effectively inform engineers of the latency of PTNNs, model zoos should report FLOP counts alongside accuracy. Also of interest is the energy usage of these PTNNs, another important property for edge devices. 
The lack of reporting of these properties may make choosing PTNNs more difficult. We recommend future works create new metrics to measure the reliability and quality of PTNNs from model zoos and develop tools for automatically measuring these properties. Publishing updated results frequently can support easier decision-making of models for deployment. 

\noindent\textbf{Naming Conventions.} The differences in the architectures of PTNNs may indicate an underlying improper documentation standard and a need for improved naming conventions in model zoos. As indicated in \cref{Architecture}, Torchvision's \emph{AlexNet} did not adhere to the origin paper while still claiming to be \emph{AlexNet}. Seemingly, model zoos are advertising PTNNs labeled as well-known DNN architectures, like \emph{ResNet} and \emph{AlexNet}, but when they do this, they really mean that the PTNNs are based on the DNN architecture and are not strict implementations. This inadequate naming convention leads to a false sense of equality and thus confusion. 
We recommend the community comprehensively document PTNN naming conventions to increase cohesion among model zoos.
Likewise, we suggest future works investigate the expectations of engineers with regards to the PTNNs from model zoos to see whether they prefer exact reproductions or more accurate and lower latency PTNNs. 
The result of such a study would inform model zoo maintainers on how to best implement and train PTNNs.

\section{Conclusion}
We present an investigation of the discrepancies between 36 image classification PTNN architectures from four popular model zoos through accuracy, latency, and architecture analyses. We find several significant discrepancies among these three axes that challenge the well-established use of PTNNs from model zoos, suggesting that an engineer will receive a PTNN with different characteristics based on the model zoo. The PTNN's goal of shortening model deployment time is diminished because of the time investment needed to verify the properties of the PTNN. We discuss the importance of future works to validate the claims of model zoos, develop automated tools for measurement, and explore best practices for implementing model zoo PTNNs.

\section*{Acknowledgments}
    
    This work was supported by gifts from Google and Cisco and by NSF-OAC award \#2107230. We thank the anonymous reviewers for their careful reading of our manuscript and their many insightful comments and suggestions.
\raggedbottom
\pagebreak
\balance

\bibliographystyle{ACM-Reference-Format}
\bibliography{./bib/refs, ./bib/WenxinZotero}

\end{document}